\documentclass[conference]{IEEEtran}
\IEEEoverridecommandlockouts

\usepackage{cite}
\usepackage{amsmath,amssymb,amsfonts}
\usepackage{algorithmic}
\usepackage{graphicx}
\usepackage{textcomp}
\usepackage{xcolor}
\usepackage{algorithm}
\usepackage{multirow}
\usepackage{booktabs}
\newtheorem{defn}{Definition}[section] 
 
\usepackage{bm}

\def\BibTeX{{\rm B\kern-.05em{\sc i\kern-.025em b}\kern-.08emT\kern-.1667em\lower.7ex\hbox{E}\kern-.125emX}}

\begin{document}

\title{AHINE: Adaptive Heterogeneous Information Network Embedding}

\author{\IEEEauthorblockN{Yucheng Lin}
\IEEEauthorblockA{\textit{AI Labs, Didi Chuxing}\\
linyucheng@didiglobal.com}
\and
\IEEEauthorblockN{Xiaoqing Yang}
\IEEEauthorblockA{\textit{AI Labs, Didi Chuxing}\\
yangxiaoqing@didiglobal.com}
\and
\IEEEauthorblockN{Zang Li}
\IEEEauthorblockA{\textit{AI Labs, Didi Chuxing}\\
lizang@didiglobal.com}
\and
\IEEEauthorblockN{Jieping Ye}
\IEEEauthorblockA{\textit{AI Labs, Didi Chuxing}\\
yejieping@didiglobal.com}
}
% \author{Yucheng Lin}
% \affiliation{%
%   \institution{AI Labs, Didi Chuxing}}
% \email{linyucheng@didiglobal.com}

% \author{Xiaoqing Yang}
% \affiliation{%
%   \institution{AI Labs, Didi Chuxing}}
% \email{yangxiaoqing@didiglobal.com}

% \author{Zang Li}
% \affiliation{%
%   \institution{AI Labs, Didi Chuxing}}
% \email{lizang@didiglobal.com}

% \author{Jieping Ye}
% \affiliation{%
%   \institution{AI Labs, Didi Chuxing}}
% \email{yejieping@didiglobal.com}

\maketitle

\begin{abstract}
Network embedding is an effective way to solve the network analytics problems such as node classification, link prediction, etc. It represents network elements using low dimensional vectors such that the graph structural information and properties are maximumly preserved. Many prior works focused on embeddings for networks with the same type of edges or vertices, while some works tried to generate embeddings for heterogeneous network using mechanisms like specially designed meta paths. In this paper, we propose two novel algorithms, GHINE (General Heterogeneous Information Network Embedding) and AHINE (Adaptive Heterogeneous Information Network Embedding), to compute distributed representations for elements in heterogeneous networks. Specially, AHINE uses an adaptive deep model to learn network embeddings that maximizes the likelihood of preserving the relationship chains between non-adjacent nodes. We apply our embeddings to a large network of points of interest (POIs) and achieve superior accuracy on some prediction problems on a ride-hailing platform. In addition, we show that AHINE outperforms state-of-the-art methods on a set of learning tasks on public datasets, including node labelling and similarity ranking in bibliographic networks.
\end{abstract}

\begin{IEEEkeywords}
Network embedding, heterogeneous information network, deep learning
\end{IEEEkeywords}

\section{Introduction}
Understanding and encoding interconnected relationships between the various elements in a network is important for solving many real-world problems, such as fraud detection \cite{ZhouHYF18}, ride-hailing demand forecasting \cite{demandForcasting}, search ranking \cite{airbnb}.  To learn feature representations for nodes and edges, many approaches such as \cite{Cao2015GraRep}, \cite{DeepWalk}, \cite{LINE}, \cite{node2vec} were proposed in the past few years, aiming to embed network elements into a low-dimensional space. A graph embedding system can be viewed as a mapping system, where the input is a graph with a set of vertices and edges, and the output are vectors for vertices.

However, most existing methods such as  \cite{Cao2015GraRep}, \cite{DeepWalk}, \cite{LINE}, \cite{node2vec}, \cite{Wang2016Structural}, \cite{Zhang2017Homophily} are for embedding homogeneous network, i.e., they can deal only with networks containing nodes and edges of one type. 
As there are many different types of nodes and relations in heterogeneous information networks (HINs) such as DBLP \cite{dblp}, YAGO \cite{Yago}, DBpedia \cite{DBpedia} and Freebase \cite{Freebase}, network embedding algorithms that capture these semantics become more important.  HINs are also widely used in industries, such as ride-hailing \cite{demandForcasting}, accommodation \cite{airbnb}, etc.  Previous homogeneous methods are not applicable for generating embeddings for heterogeneous networks.  

Relations between nodes in a HIN are much more complex than those in homogeneous networks. The proximity among nodes is not just a measure of closeness or distance, but also some other semantics (e.g., type of relations between author and paper, type of relations between co-authors in DBLP). Thus embedding models based on node closeness are not suitable for HINs. There are several meta path based methods that try to explore and maintain the rich semantic and structural information in HINs, such as \cite{Ji2018Attention}, \cite{MAP}, \cite{metapath2vec}, \cite{Wang2018Unsupervised}, \cite{Wei2018SHINE}, \cite{Wang2017Distant}, \cite{Shi2017PReP}, \cite{Ming2016Text}, \cite{hhne}. The meta path based methods often require experts to specify the meta paths or provide supervision to select meta paths. Furthermore, the set of meta paths often reflects part of the semantic meanings in the HINs as experts only supply part of patterns of relationships in the HINs. Other kinds of semantic relations which are not in the set of meta paths cannot be captured by these methods.  

% There are other methods that deal with HIN embeddings without the usage of meta paths. \cite{Shi2018Easing} proposed a two-stage HEER architecture for learning HIN embeddings via learning edge representations. But it only reflects the relationship between adjacent nodes in HINs. 

There are other methods that deal with HIN embeddings without the usage of meta paths, such as network partition based methods \cite{herec} \cite{eoe}, neural network based methods \cite{hne} \cite{bl-mne} \cite{shine}. These methods are usually only applicable to specific tasks or networks and are not well suited for use in general scenarios.

To cope with the challenges of HIN embedding, we propose two novel methods, General Heterogeneous Information Network Embedding (GHINE) and Adaptive Heterogeneous Information Network Embedding (AHINE). By modeling relation types as deep neural layers, the two algorithms are able to transform nodes in a heterogeneous network into distributed representations while preserving the semantic proximities between nodes. We summarize our major contributions as follows:

(1) We propose two unsupervised learning algorithms, GHINE and AHINE, for generating HIN embeddings that preserve complex relationships between nodes without any knowledge about meta paths or predefined rules. 

(2) The edge-based algorithm (GHINE) and chain-based algorithm (AHINE) are presented to encode first order and higher order semantic relations among nodes. Different types of edges are modeled by different deep neural networks .

(3) Experiments based on HIN datasets demonstrate the effectiveness of AHINE and GHINE. They outperform state-of-the-art methods in tasks on two public HINs. We also apply AHINE on a real-world large-scale ride-hailing dataset collected in Beijing to get embeddings for POI grid cells. The embeddings help greatly improve the ride-hailing activity prediction service.

\section{Related Work}
\subsection{Homogeneous Network Embedding}
Network embedding has attracted extensive attention recently. Inspired by word2vec \cite{skipgram}, DeepWalk proposed in \cite{DeepWalk} generates truncated random walks and treats these walks as the equivalent of sentences. By applying Skip-gram model to the ``sentences'', latent representations of nodes are well extracted. Node2vec \cite{node2vec} further extended DeepWalk by designing a flexible objective function and providing parameters $p$, $q$ to tune the explored search space. LINE \cite{LINE} is another widely used network embedding model, which learns feature representations in two separate phases. In the first phase, it learns first-order embedding by simulation over immediate neighbors of nodes. In the second phase, it learns second-order embedding by sampling nodes at a $2$-hop distance. Struc2vec \cite{struc2vec} focuses on capturing structural equivalence between nodes.

Over the last few years, there has been a surge of Graph Neural Networks (GNNs) studies \cite{gnnsurvey}, which operate deep learning based methods on graph domain, such as Graph Convolutional Networks (GCNs) \cite{gcn}, GraphSage \cite{graphsage}, Graph Attention Networks (GATs) \cite{gat}, etc. Though they show high interpretability and achieve convincing performance, most GNNs still seem to have serveral limitations: (1) only focus on homogeneous networks (2) supervised labels are required (3) hard to deal with large-scale graphs due to high memory requirement.

\subsection{Heterogeneous Network Embedding}

The aforementioned methods focus on homogeneous information networks, where all the nodes are of the same type and all the edges share the same relation. In recent years, a few works have been done to learn latent representations from heterogeneous information networks (HINs).

A main class of embedding studies for HINs are based on meta paths proximities. Take bibliographic datasets for example. Meta paths like ``Author-Paper-Author'' (``APA'') or ``Author-Paper-Venue-Paper-Author'' (``APVPA'') are generated first, which preserve the information of coauthor or similar research field. ESim \cite{shang2016meta} accepts user-defined meta paths as guidance to learn vertex vectors in a user-preferred embedding space. Metapath2vec \cite{metapath2vec} formalizes random walks on a meta path scheme like ``APVPA'', and performs the Skip-gram model on it for HIN embedding. Metapath2vec++ \cite{metapath2vec} proposes an advanced Skip-gram framework, in which the softmax function is normalized with respect to the node type of the context. HIN2Vec \cite{fu2017hin2vec} learns HIN embeddings by conducting multiple prediction training tasks jointly to learn representation from meta paths. HHNE \cite{hhne} measures the node proximity in hyperbolic spaces instead of Euclidean. 

However, most of these models show poor generality, as meta paths are required to be ready in advance. In many practical applications, users have to design different meta paths to capture different graph semantics. This makes these methods difficult to apply widely.

In addiction to meta path based methods, some methods such as HERec \cite{herec} and EOE \cite{eoe} decompose the HIN into sub-networks and optimize node proximity within each sub-network. Some other works are inspired by deep learning technology and learn non-linear mapping functions for HIN embedding by training neural networks, such as HNE \cite{hne}, BL-MNE \cite{bl-mne}, SHINE \cite{shine}, etc. These methods usually target at some specific tasks and are hard to handle general cases.

\section{Preliminaries} 
\begin{defn}
An $\bf{information}$  $\bf{network}$ is a directed graph $G = (V, E)$ with a node type mapping function $\phi: V \rightarrow \Gamma$ and an edge type mapping $\psi: E \rightarrow \Omega$, where each node $v \in V$ belongs to a node type $\phi(v) \in \Gamma$, and each edge $r \in E$ belongs to an edge type $\psi(r) \in \Omega$. 
\end{defn}

\begin{defn}
An information network $G = (V, E)$ is a $\bf{heterogeneous}$ $\bf{information}$ $\bf{network}$ $\bf{(HIN)}$ if 
$\phi(v_1) \neq \phi(v_2)$, $ \ \exists$ $v_1, v_2 \in V$ or $\psi(r_1) \neq \psi(r_2)$,  $\ \exists$ $r_1, r_2 \in E$. 
\end{defn}

\begin{figure}[ht]
  \centering
  \includegraphics[width=0.7\linewidth]{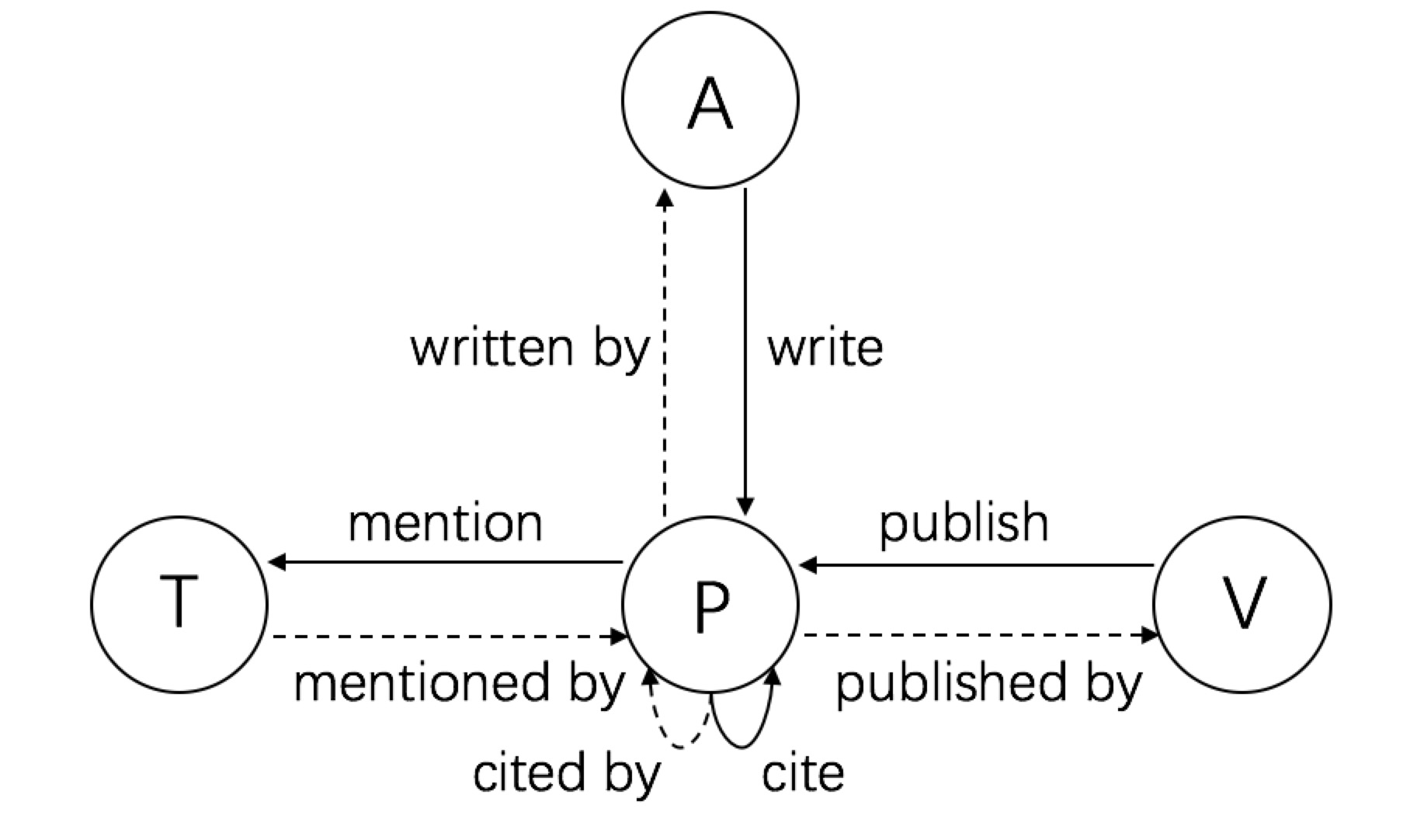}
  \caption{Schemas of DBLP Heterogeneous Information Network}
  \label{fig:dblp}
\end{figure}

Fig.~\ref{fig:dblp} shows the schema of DBLP HIN, where nodes A, T, P, V correspond to author, topic, paper and venue, respectively. There are also different kinds of relations between nodes, such as ``publish'', ``write'', ``cited by'', etc. 

\begin{defn}
A $\bf{meta}$ $\bf{path}$ $P$ is an ordered list of node types  $\gamma_1$, $\gamma_2$, ..., $\gamma_n$ connected by edge types $e_1$, $e_2$, ..., $e_{n-1}$ as follows:
$$P=\gamma_1 \stackrel{e_1}{\longrightarrow} \gamma_2 \cdots \gamma_{n-1} \stackrel{e_{n-1}}{\longrightarrow} \gamma_n.$$
\end{defn}

\begin{figure}[ht]
  \centering
  \includegraphics[width=0.8\linewidth]{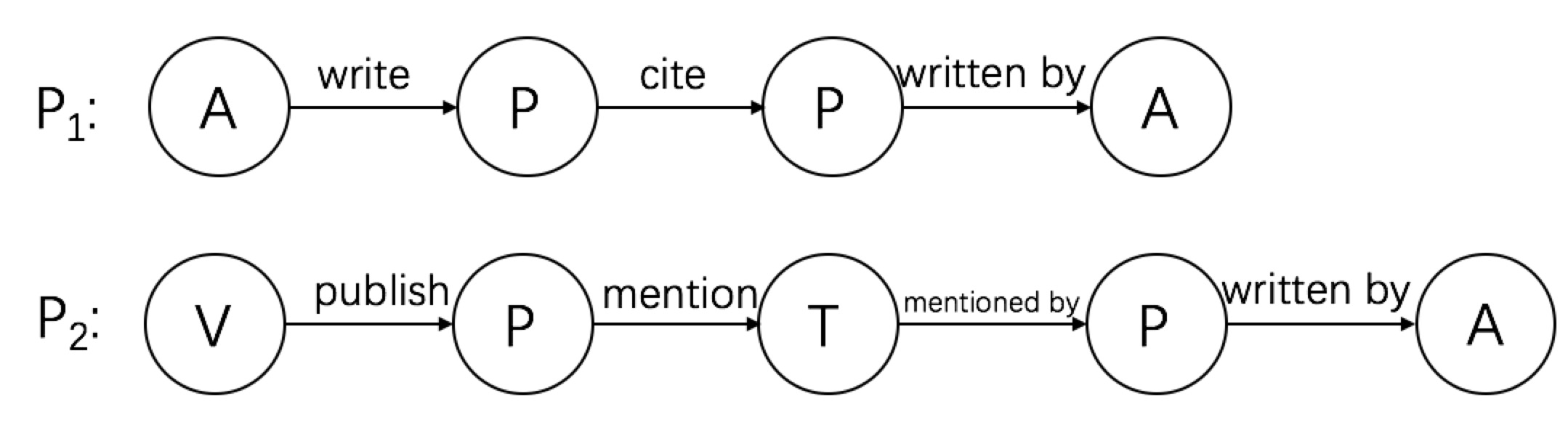}
  \caption{Meta path examples for DBLP HIN in Fig.~\ref{fig:dblp} }
  \label{fig:meta_path}
\end{figure}

An instance of the meta path $P$ is a real path in the HIN with the pattern of $P$.  Fig. ~\ref{fig:meta_path} illustrates two examples of meta paths for DBLP HIN shown in Fig.~\ref{fig:dblp}.

\begin{defn}
$\bf{Heterogeneous}$ $\bf{Information}$ $\bf{Network}$ $\bf{Embedding}$: Given a heterogeneous information network $G$, find a d-dimensional representation $\Phi \in R^{|V|*d}$, $d << |V|$ that is able to explore and maintain the semantic and structural relations among them. 
\end{defn}

\section{Methodology}
In this section, we introduce our methodology for embedding HINs. We first discuss how to model the heterogeneous information network embedding problem in a general way in Section ~\ref{subsection:ghine}.  Then, we introduce an adaptive version which is called AHINE in Section ~\ref{subsection:ahine}. Finally, we present an example from the ride-hailing application in Section ~\ref{subsection:didi}.

\subsection{General HINE}\label{subsection:ghine}
An effective embedding learning method should consider the differences between different relationships. GHINE ($\bf{General}$  $\bf{Heterogenous}$ $\bf{Information}$  $\bf{Network}$  $\bf{Embedding}$) treats each relationship as a non-linear function which is formulated by a deep learning model. For example, if there exists an edge type $e$ from node $v_1$ to node $v_2$, we would like to formulate it as follows:

\begin{equation}
f_e(\Phi(v_1)) = \Phi(v_2),
\end{equation}

where $f_e$ is the function for edge type e, $\Phi(v_1)$ is GHINE for $v_1$ and $\Phi(v_2)$ is GHINE for $v_2$. 

Like DeepWalk \cite{DeepWalk} does, our algorithms extend Skip-gram \cite{skipgram} architecture to networks. In DeepWalk, it constrains a node’s embedding to be similar to its context nodes in random walk by calculating their similarities. However such method only encodes network closeness but does not take into account relation information. Thus, we define the similarity between adjacent nodes by considering the relation types. If there is an edge type $e$ from node $v_i$ to $v_j$, we define the similarity between $v_i$ and $v_j$ via relation $e$ as 

\begin{equation}
s_e(v_i, v_j) = f_e(\Phi(v_i))^{T} \cdot \Phi(v_j),
\end{equation}

where $\Phi(v_i)$ (or $\Phi(v_j)$) $\in R^d$ is the embedding of node $v_i$ (or $v_j$). The probability of $Pr(v_j|v_i, e)$ is modeled via softmax: 

\begin{equation}
Pr(v_j | v_i, e) = \frac{e^{s_e(v_i, v_j)}}{\sum_{v'\in V}e^{s_e(v_i, v')}}.
\end{equation}

To learn node embeddings, the GHINE algorithm first generates a set of triples ($v_i$, e, $v_j$) by random edge sampling on HIN. After that, stochastic gradient descent is used to learn the parameters of $\Phi$ and $f_e$. At each iteration, a set of triples with the same edge type as a mini batch is processed to update the gradients to minimize the following objective:

\begin{equation}
L_{ij} = - logPr(v_j | v_i, e).
\end{equation}

We use negative sampling to approximate the objective function in order to speed up training process. Formally, parameters of $\Phi$ and $f_e$ are updated as follows:
\begin{equation}
\label{equ:phi}
\Phi = \Phi - \eta\frac{\partial L_{ij}}{\partial \Phi},
\end{equation}
\begin{equation}
\label{equ:f_e}
f_e = f_e - \eta\frac{\partial L_{ij}}{\partial f_e},
\end{equation}

where $\eta$ is the learning rate. 

\begin{algorithm}[htb]
\caption{THE GHINE ALGORITHM}
\label{alg:ghine}
\begin{algorithmic}[1]
\REQUIRE

(1) A heterogeneous information network: $G = (V, E)$;

(2) Maximum number of iterations: MaxIterations;

(3) mini batch size: $b$;

(4) learning rate: $\eta$;

(5) Edge type list: $L$.
\ENSURE

Node embedding $\Phi(\cdot)$ for each $v \in V$

\STATE Initialize $|L|$ neural networks with the same input / output dimension
\STATE $S \leftarrow$ generate a set of triples ($v_i$, $e_k$, $v_j$) according to $G$ where $e_k \in L$
\STATE Iterations $\leftarrow 0$
\REPEAT
\STATE get a mini batch of size $b$ of ($v_i$, $e_k$, $v_j$) from S with the same edge type $e_k$;
\STATE update parameters of $\Phi$ and $f_{e_k}$ by Eq.~\ref{equ:phi} and Eq.~\ref{equ:f_e}
\STATE Iterations $\leftarrow$ Iterations + $1$
\UNTIL Iterations $\geqslant$ MaxIterations or convergence
\RETURN $\Phi$

\end{algorithmic}
\end{algorithm}

In GHINE, the layers of embedding and softmax share the same weights. Each type of edge has a different deep network layer to represent. The only constraint for these layers is that the number of dimensions of the input and output layers of the deep networks must be the same as the number of dimensions of node embeddings. During the training process, it selects the proper DNN layers for the specific relation. 

\begin{figure}[ht]
  \centering
  \includegraphics[width=\linewidth]{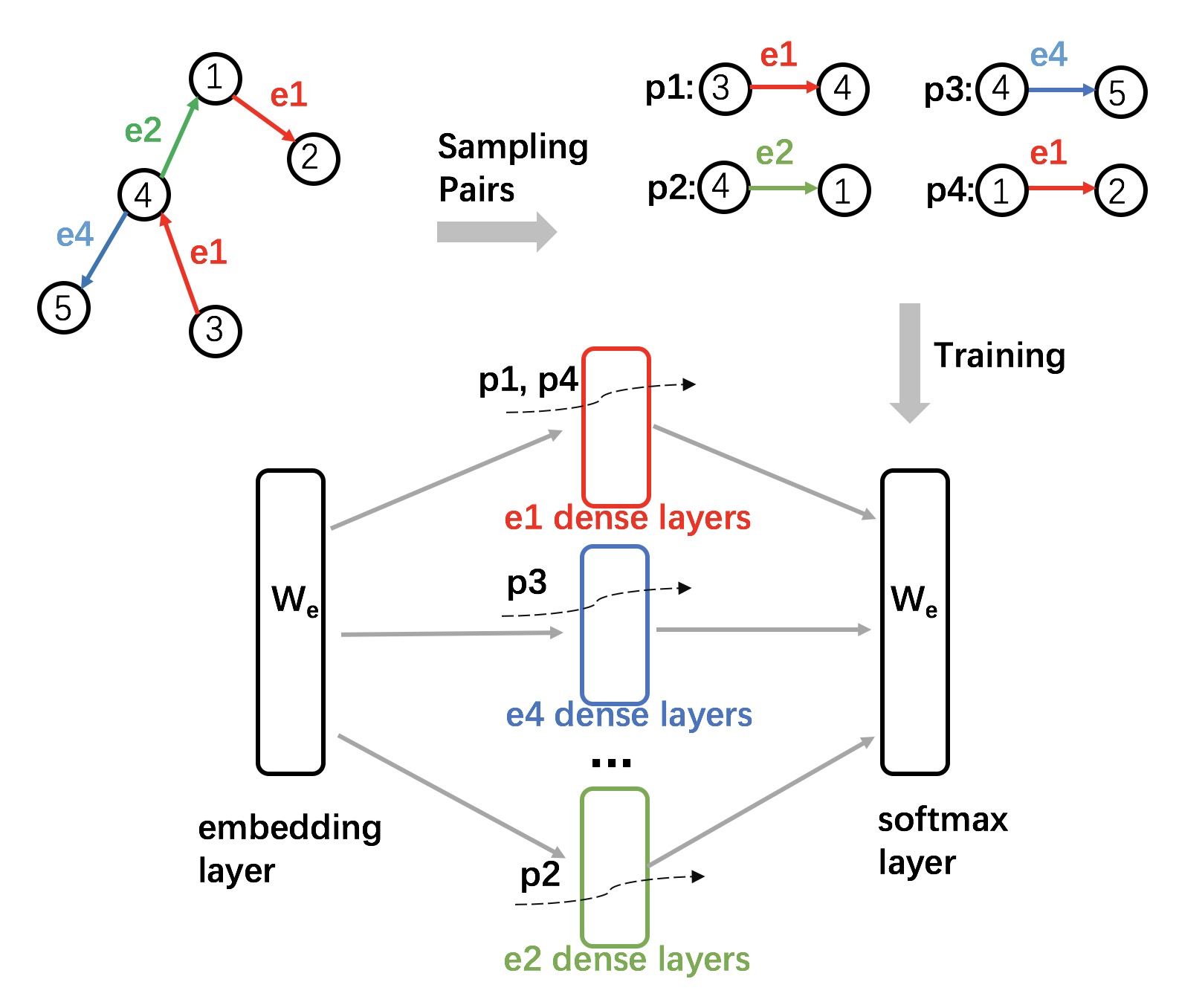}
  \caption{General Heterogeneous Information Network Embedding}
  \label{fig:ghine}
\end{figure}

The key steps of GHINE are described in Algorithm~\ref{alg:ghine}. Fig.~\ref{fig:ghine} gives an example of GHINE. Four triples $p_1$ ($v_3$ $\stackrel{e_1}{\longrightarrow}$ $v_4$), $p_2$ ($v_4$ $\stackrel{e_2}{\longrightarrow}$ $v_1$), $p_3$ ($v_4$ $\stackrel{e_4}{\longrightarrow}$ $v_5$), and $p_4$ ($v_1$ $\stackrel{e_1}{\longrightarrow}$ $v_2$) are sampled from a HIN. $e_1$, $e_2$, $e_4$ are types of edges. Different types of edges are colored differently. GHINE models each edge type using a distinct DNN. When processing the training samples $p_1$ or $p_4$, the weights of the embedding/softmax layers and DNN layers corresponding to type $e_1$ will be updated.

\subsection{Adaptive HINE}\label{subsection:ahine}
Furthermore, we have proposed a more flexible model, called $\bf{Adaptive}$ $\bf{Heterogenous}$ $\bf{Information}$ $\bf{Network}$ $\bf{Embedding}$ (AHINE). AHINE tries to encode the relationships between nodes with distance $> 1$ in the HIN. In order to achieve this, relation chains (of length $\geq$ $1$) are generated as training examples by random walks or by designing meta paths. These training samples preserve the relationship between both adjacent and non-adjacent nodes. In this sense, GHINE is a special case of AHINE, i.e., the length of each relation chain is always $1$.

In this method, as each input training example may have different lengths and different types of relations, the computation graph will change accordingly. For every input chain, the algorithm tries to predict the last node based on the first node and the information carried by the relation chains between them.

Suppose there is a relation chain $v_i$ $\stackrel{e_1}{\longrightarrow}$ ... $\stackrel{e_{m}}{\longrightarrow}$ $v_j$, AHINE tries to predict $v_j$'s embedding by a neural network $f_{e_{m}}(...(f_{e_1})...)$ as
\begin{equation}
f_{e_m}(f_{e_{m-1}}(....f_{e_1}(\Phi(v_i))...)) = \Phi(v_j).
\end{equation}

The proximity between $v_i$ and $v_j$ via a relation chain ($e_1$, $e_2$, ... $e_m$) can be modeled as:

\begin{equation}
s_{e_1, e_2,...e_m}(v_i, v_j) = f_{e_m}(f_{e_{m-1}}(...f_{e_1}(\Phi(v_i))...))^{T} \cdot \Phi(v_j).
\end{equation}

Then the probability $Pr(v_j | v_i, e_1, e_2, ..., e_m)$ is:

\begin{equation}
Pr(v_j | v_i, e_1, e_2, ..., e_m) = \frac{e^{s_{e_1, e_2, ... e_m}(v_i, v_j)}}{\sum_{v'\in V}e^{s_{e_1, e_2, ... e_m}(v_i, v')}}.
\end{equation}

The AHINE algorithm generates samples like ($v_i$, $e_1$, ...$e_m$, $v_j$) by random walks or meta path patterns. During the training process, the structure of network model is determined by the relation chain between $v_i$ and $v_j$. This mechanism allows us to feed any kind of meta paths into the model. Stochastic gradient descent is used to learn the parameters of $\Phi$ and $f_e$ . At each iteration, a set of samples with the same relation chains is processed as a mini batch to update the gradients to minimize the following objective,

\begin{equation}
L'_{ij} = -log Pr(v_j | v_i, e_1, e_2, ...., e_m) .
\end{equation}

Same as GHINE, negative sampling is used to speed up training process. Parameters $\Phi$ and $f_{e_k}, k \in [1, m]$ are updated as follows:

\begin{equation}
\label{equ:ahine_phi}
\Phi = \Phi - \eta\frac{\partial L'_{ij}}{\partial \Phi},
\end{equation}

\begin{equation}
\label{equ:ahine_f_e}
f_{e_k} = f_{e_k} - \eta\frac{\partial L'_{ij}}{\partial f_{e_k}} ,    k \in [1,m].
\end{equation}

In practice, we use the samples where the length of edges is $1$ to initialize dense layers and embeddings for nodes first. In other words, GHINE is used to initialize dense layers and embeddings for node. And then,  the samples where the length of edges is more than $1$ are then used to optimize the total network weights. 

\begin{figure}[ht]
  \centering
  \includegraphics[width=0.9\linewidth]{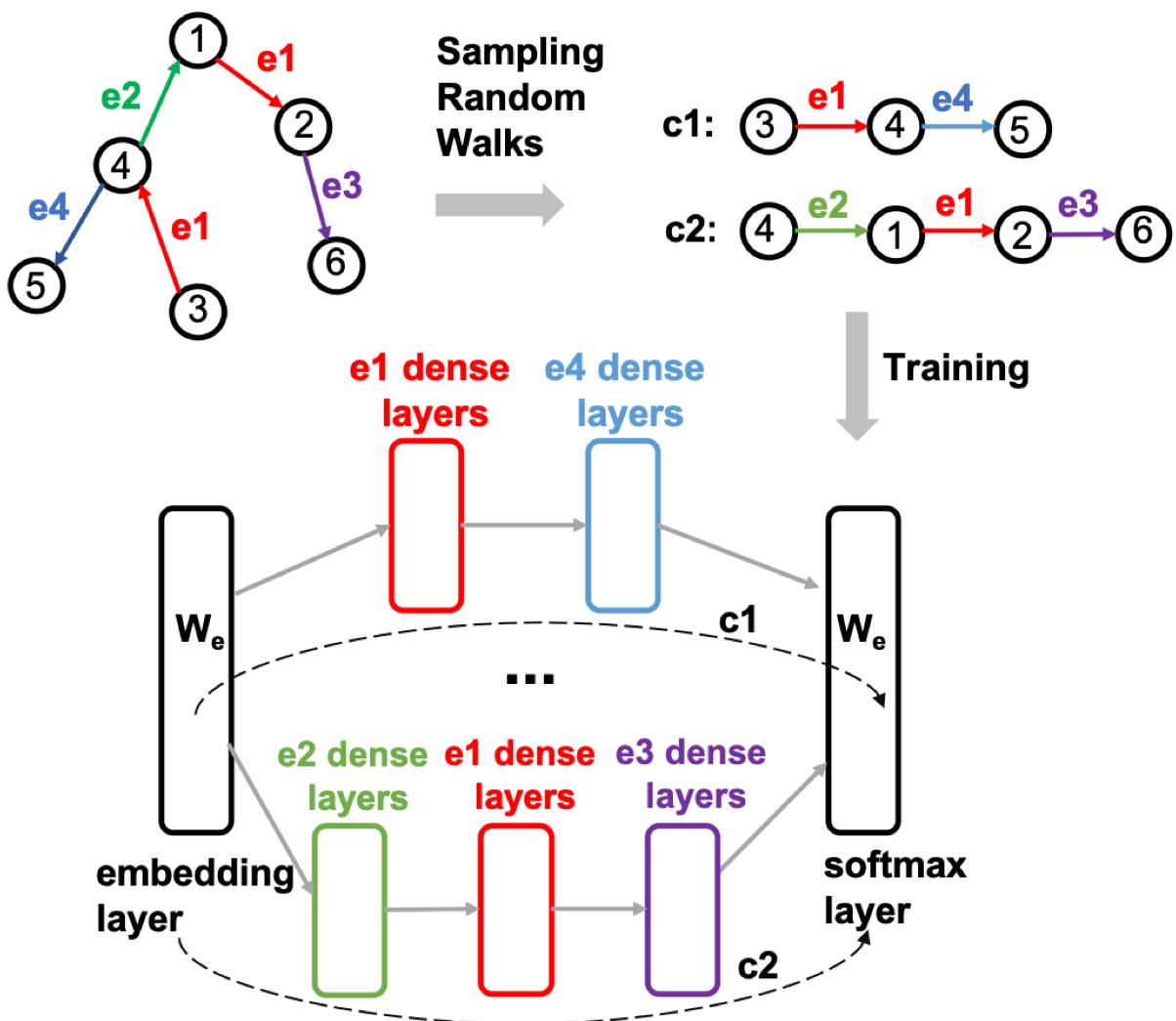}
  \caption{Adaptive Heterogeneous Information Network Embedding}
  \label{fig:ahine}
\end{figure}

Fig. \ref{fig:ahine} follows the same example as above. Two chains $c_1$ ($v_3$ $\stackrel{e_1}{\longrightarrow}$ $v_4$ $\stackrel{e_4}{\longrightarrow}$ $v_5$) and $c_2$ ($v_4$ $\stackrel{e_2}{\longrightarrow}$ $v_1$ $\stackrel{e_1}{\longrightarrow}$ $v_2$ $\stackrel{e_3}{\longrightarrow}$ $v_6$) are sampled from the HIN by random walks. For training example $c_1$, AHINE tries to predict $v_5$ by giving $v_3$'s node embedding. When processing $c_1$, the weights of the embedding/softmax layers and DNN layers corresponding to type $e_1$ and $e_4$ will be updated.

Algorithm~\ref{alg:dcg} describes the details of how to construct dynamic computation graph for each relation chain. The main steps of AHINE are shown in Algorithm~\ref{alg:ahine}. 

\begin{algorithm}[htb]
\caption{CONSTRUCTION FOR DYNAMIC COMPUTATION GRAPH ALGORITHM}
\label{alg:dcg}
\begin{algorithmic}[1]
\REQUIRE

(1) Edge type list: $L$;

(2) Max chain length: $c$;

(3) $|L|$ neural networks: one neural network for each edge type 

\ENSURE

dynamic computation graph set $S$

\STATE Computation graph set $S$ $\leftarrow$ $\phi$
\STATE $chain\_length$ $\leftarrow 1$
\REPEAT
\STATE $P$ $\leftarrow$ get all relation chains with length $chain\_length$
\STATE $p_{ind}$ $\leftarrow 0$
\REPEAT
\STATE $g$ $\leftarrow$ construct computation graph according to the sequence of $P[p_{ind}]$
\STATE add $g$ to $S$
\STATE $p_{ind}$ $\leftarrow$ $p_{ind} + 1$
\UNTIL $p_{ind}$ $\geqslant$ $|P|$
\STATE $chain\_length$ $\leftarrow$ $chain\_length + 1$
\UNTIL $chain\_length > c$
\RETURN $S$

\end{algorithmic}
\end{algorithm}

\begin{algorithm}[htb]
\caption{THE AHINE ALGORITHM}
\label{alg:ahine}
\begin{algorithmic}[1]
\REQUIRE

(1) A heterogeneous information network: $G = (V, E)$;

(2) Maximum number of iterations: MaxIterations;

(3) mini batch size: $b$;

(4) learning rate: $\eta$;

(5) Edge list: $L$;

(6) Max chain length: $c$.
\ENSURE

Node embedding $\Phi(\cdot)$ for each $v \in V$

\STATE Initialize $|L|$ neural networks and $\Phi$ by GHINE described in Algorithm~\ref{alg:ghine}
\STATE construct a dynamic computation graph set by $|L|$ neural networks with chain length $\leqslant$ c  as described in Algorithm~\ref{alg:dcg}
\STATE S $\leftarrow$ generate a set of samples ($v_i$, $e_1$...$e_k$...$e_m$, $v_j$) according to G where $e_k \in L$, $1\leqslant k \leqslant c$ and $m \leqslant c$
\STATE Iterations $\leftarrow 0$
\REPEAT
\STATE get a mini batch of size $b$ of ($v_i$, $e_1$, ...$e_k$, ...$e_m$, $v_j$) from $S$ with the same relation list [$e_1$, ...$e_k$, ...$e_m$];
\STATE update parameters of $\Phi$ and $f_{e_k}$ by Eq.~\ref{equ:ahine_phi} and Eq.~\ref{equ:ahine_f_e}
\STATE Iterations $\leftarrow$ Iterations + 1
\UNTIL Iterations $\geqslant$ MaxIterations or convergence
\RETURN $\Phi$

\end{algorithmic}
\end{algorithm}

\subsection{AHINE in Ride-hailing Platform}\label{subsection:didi}
We carry out experiments in a large-scale ride-hailing platform.  We built a large graph where each node represents one of the $28,929$ square grid cells in Beijing.  We aim to learn a low-dimensional representation for each grid cell, capturing intuitive aspects of locations such as residential areas which have different characteristics from office buildings. In addition, we constrain areas with similar functions to have similar features, and adjacent areas to be close in embedding space. These aspects will allow us to improve other machine learning tasks involving POIs.

Our data consists of passenger ride orders.  Each order is a triple ($POI_i$, $rel_t$, $POI_j$), where $POI_i$ is the source cell, $POI_j$ is the destination cell, and $rel_t$ is one of $10$ discrete time values.  Time values are members of the cross-product:

\begin{equation}
\left\{
\begin{aligned}
peak \  morning \\
day \  time \\
peak \  evening \\
dusk \  to \  midnight \\
midnight \  to \  morning 
\end{aligned}
\right\}
*
\left\{
\begin{aligned}
weekday \\
weekend 
\end{aligned}
\right\}.
\end{equation}

\begin{figure}[ht]
  \centering
  \includegraphics[width=0.8\linewidth]{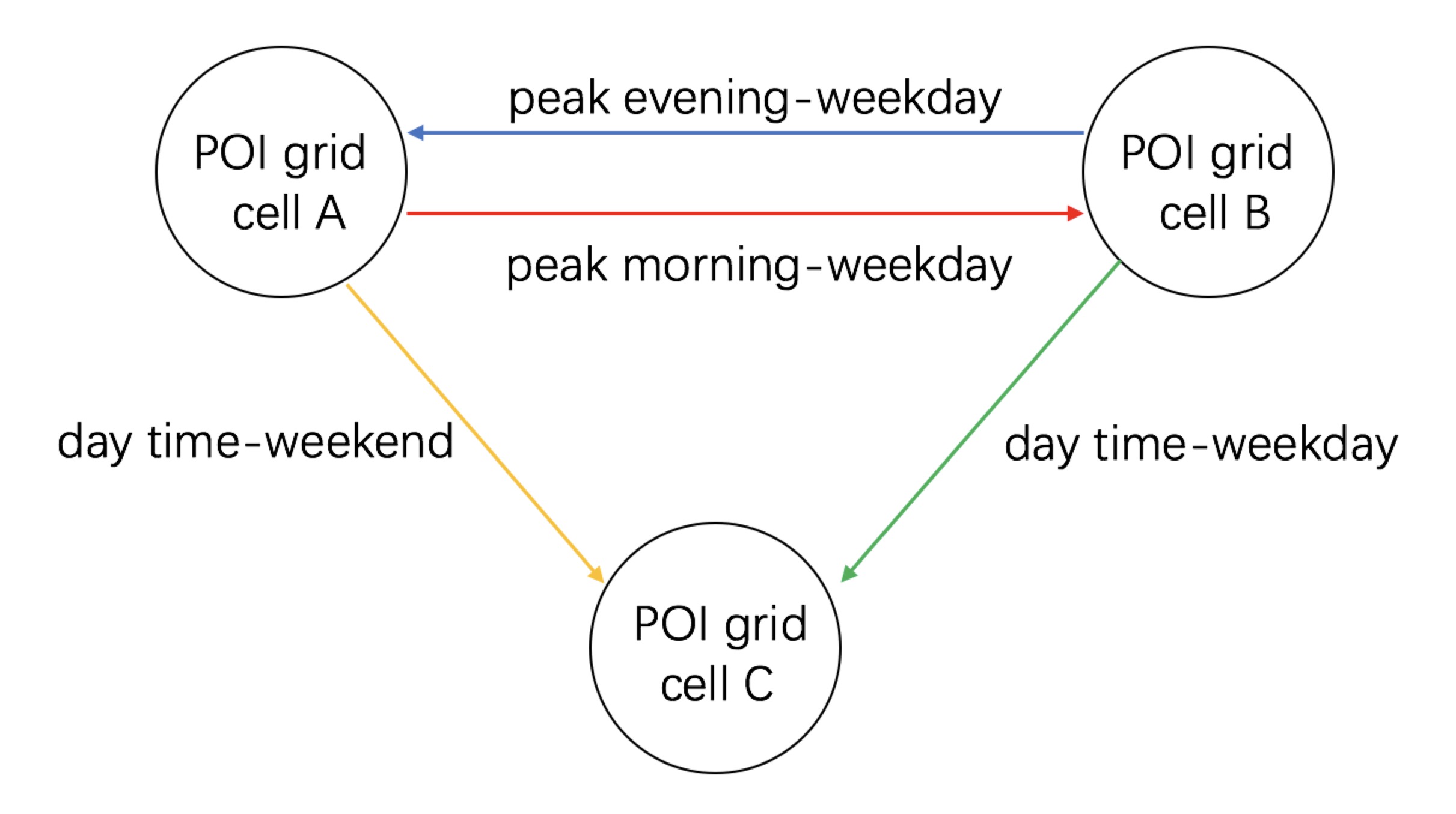}
  \caption{Ride-hailing POI Grid Heterogeneous Information Network}
  \label{fig:poi_grid}
\end{figure}

For each order ($POI_i$, $rel_t$, $POI_j$), we add to our graph $G$ an edge of type $rel_t$ from $v_i$ to $v_j$.  We then run AHINE on this graph, which contains $14,804,324$ edges. Fig.~\ref{fig:poi_grid} illustrates an example for ride-hailing POI grid heterogeneous information network. In this example, some passenger calls a taxi from POI grid cell A to POI grid cell B in the peak morning of weekday, while another person generates an order from POI grid cell B to POI grid cell A in the peak evening of weekday.  We generate samples such as  ``POI grid cell A $\stackrel{rel_1}{\longrightarrow}$ POI grid cell B'', ``POI grid cell B $\stackrel{rel_2}{\longrightarrow}$ POI grid cell A'' where $rel_1$ means peak morning in weekday and $rel_2$ means peak evening in weekday. These kinds of samples can be used in GHINE to obtain embeddings for each POI grid cell. 

If we would like to emphasize a continuous trip by the same passenger, we propose to use training samples such as ``POI grid cell A $\stackrel{rel_1}{\longrightarrow}$ POI grid cell B $\stackrel{rel_2}{\longrightarrow}$ POI grid cell C $\stackrel{rel_3}{\longrightarrow}$ POI grid cell A'' where $rel_1$ means peak morning in weekday, $rel_2$ means daytime in weekday and $rel_3$ means peak evening in weekday. To capture the information contained in this example, the training model selects the dense layers for relation ``peak morning-weekday'', ``daytime-weekday'' and ``peak evening-weekday'' in sequence with both the input and output as POI grid cell A.

\section{Experiments}
In this section, we demonstrate the effectiveness of the presented AHINE frameworks for HIN representation learning. We first introduce three heterogenous network
 datasets in Section ~\ref{subsection:datasets}. Then, in Section ~\ref{subsection:setup}, we introduce the experimental setup and competing algorithms. In Section ~\ref{subsection:didi_res} and Section ~\ref{subsection:pub_res}, experimental results and analysis are presented. 

\subsection{Datasets}\label{subsection:datasets}
Three heterogenous network datasets are used in our experiments, including a ride-hailing dataset from a real-world platform and two public bibliographic datasets.

\textbf{RH} Ride-Hailing (RH) dataset is a large, directed, multi-edged HIN, generated from ride-hailing records in Beijing. This graph is composed of $28,929$ nodes and $14,804,324$ edges of $10$ types in total. Each node represents a POI grid cell and each edge represents a ride-hailing record. Detailed information can be found in Section ~\ref{subsection:didi}.

\textbf{DBIS and AMINER} We also conduct our experiments on two public bibliographic datasets, including the Database and Information Systems (DBIS) dataset \cite{dbis} and the Aminer Computer Science (AMINER) dataset \cite{aminer}. DBIS, a subset of DBLP dataset \cite{dblp}, was constructed by Sun et al. \cite{dbis}. It contains all $464$ venues in DBLP and corresponding $60,694$ authors and $72,902$ publications. AMINER consists of $1,693,531$ computer scientists and $3,194,405$ papers from $3,883$ computer science venues. Both datasets are HINs with three types of nodes (Author, Paper, Venue) and four corresponding relationships (``write'', ``written by'', ``publish'', ``published by'') among them.

\subsection{Experimental Setup}\label{subsection:setup}

Traveling graphs and bibliographic graphs share significant distinctions in many aspects. For example, it's difficult to design meta paths for the RH data. Thus, we compare GHINE and AHINE with distinct baseline methods and settings. 

On RH dataset, DailyWalk, DeepWalk \cite{DeepWalk}, LINE \cite{LINE}, xNetMF \cite{regal}, struc2vec \cite{struc2vec} are implemented for the purpose of comparison. All embeddings are in the same dimension of $dim=30$. 

It is worth noting that the main idea of our DailyWalk model is that there are some hidden connections in all places that a person has been to for a period of time. Instead of using random walks or meta paths, DailyWalk generates walks by sequentially connecting the POIs appeared in real-world daily ride-hailing records. For example, suppose a passenger travels from $POI_1$ to $POI_2$, and then moves to $POI_3$ from $POI_2$ in the same day, we get a walk ``$POI_1{\rightarrow}POI_2{\rightarrow}POI_3$''. Similar to DeepWalk, we learned embedded representations by applying Skip-gram to the generated daily walks. 

On DBIS and AMINER data, DeepWalk, LINE, struc2vec, metapath2vec, metapath2vec++ \cite{metapath2vec}, HHNE \cite{hhne} are included for comparison. All the embeddings share the same dimension of $dim=50$. 

For all the walk based models (DeepWalk, struc2vec, metapath2vec, metapath2vec++, HHNE), we use the same parameters: 

(1) The number of walks per node $w=100$; 

(2) The max length for each walk $l=50$; 

(3) The context neighborhood size $win=3$;

(4) The size of negative samples $neg=5$.

(5) The lower bound of node frequency $min\_count=5$

% The detailed information and settings are listed as below.

% \begin{itemize}

% \item \textbf{DeepWalk} \cite{DeepWalk} Through experiments, we find that the choice between CBOW \cite{word2vec} and Skip-Gram architectures does not yield significant differences. Meanwhile, negative sampling and hierarchical softmax technique show similar performance too. Therefore, we choose Skip-Gram architecture and negative sampling technique with the above parameters for comparison.

% \item \textbf{LINE} \cite{LINE} LINE is a method considering first-order and second-order proximities between nodes. In our experiments, we employ an advanced version of LINE, which computes the first-order and second-order embeddings separately and then concatenates them. Define $dim_{1st}$ as the length of first-order embedding and $dim_{2nd}$ as the second-order embedding's length. For each node, we choose equal representation length $dim_{1st}=dim_{2nd}$, specifically, $dim_{1st}=dim_{2nd}=15$ for RH and $dim_{1st}=dim_{2nd}=25$ for DBIS/AMINER.

% \item \textbf{metapath2vec/metapath2vec++} \cite{metapath2vec} Different from DeepWalk, metapath2vec and metapath2vec++ use meta path based walks instead of global random walks. The generated walks are fed to Skip-gram or heterogeneous Skip-gram model to perform node embedding. According to Dong et al.'s suggestion in \cite{metapath2vec} and their code implementation, we choose the meta path scheme ``APVPA'' and erase the intermediate entities ``P'' before performing Skip-gram. 

% \item \textbf{DailyWalk} 

% \end{itemize}
AHINE takes ``$node_{first}$, $rel_1$, $\cdots$, $rel_i$, $node_{last}$'' sequences as input, with max chain length $c$ and $1\le i\le c$. GHINE is a special case of AHINE that all chain lengths are always $1$. It takes ``$node_1$, $rel_1$, $node_2$'' sequences as input.

\begin{figure}[ht]
  \centering
  \includegraphics[width=0.8\linewidth]{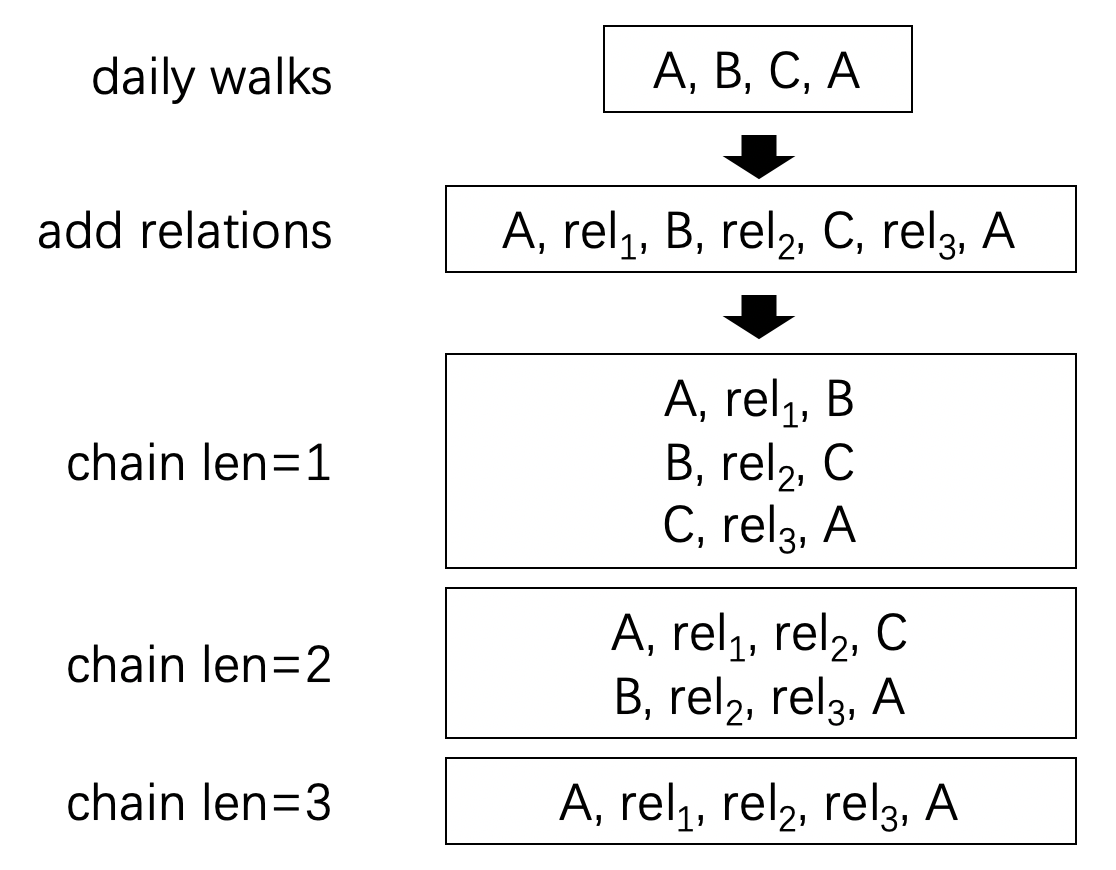}
  \caption{Generate the training samples of AHINE.}
  \label{fig:ahine-input}
\end{figure}

For RH data, Fig.~\ref{fig:ahine-input} gives an instruction of generating the training samples of AHINE. We first extract daily walks of POI grid cells from passengers' daily travel trajectories. The walks are represented in the form of a sequence in which each element of the sequence represents a POI grid cell. We then insert relations between adjacent POIs according to the order time. After that, we extracted the subsequence according to the set length. Intermediate nodes are removed for each subsequence, leaving only the edges and the first and last nodes.

Max chain length $c$ is set to $3$ in this experiment. From a daily walk including $n$ ($n\ge2$) POI nodes, we form $max(0, n-3)$ samples of chain length $3$, $max(0, n-2)$ samples of chain length $2$ and $n-1$ samples of chain length $1$ in total.

Then, we construct $10$ neural network modules for $10$ time-based relations, with the same structure and the identical input/output size, which is $30$, equal to the length of representation. There are numerous ways to construct a neural network module, where the simplest is to construct a two-layer network, without any hidden layer. Deep neural networks are worthwhile trying with adequate hardware supports. 

In our experiments, we build a $4$-layer network module for each relation, with an input layer of length $30$, an output layer of length $30$ and two hidden layer of length $200$. $ReLU$ is used as the activation function between layers and between connected neural network modules.

We take advantage of dynamic computation graph in TensorFlow \cite{abadi2016tensorflow} to combine these neural network modules dynamically. By using dynamic computation graph, it is equivalent to construct $10$ types of AHINE structure with single NN module, $100$ types of AHINE structure with double NN modules and $1,000$ types of AHINE structure with triple NN modules in total. 

With training samples prepared and the AHINE model constructed, we start the training process based on Algorithm~\ref{alg:ahine}. Training data is organized in a batch size of $32$ and average negative sampling size for each training sample is set as $neg=5$. The maximum number of iterations is set as $200$. So, we end the training process until the convergence or $epoch \ge 200$.

For the bibliographic datasets, in the similar way, we generate sequences by adding corresponding relations between adjacent nodes in random walks (or meta paths) and then erase all the nodes except for the first and the last. The walks are generated with the same parameters listed above, namely the number of walks per node $w=100$ and walk length $l=50$. Same parameters are used to construct the model, except for that relation number is $4$ and the length of input/output layer is $50$.

There are several useful tips for training AHINE. (1) Pre-training is recommended. In our tasks, we pre-train a GHINE, equal to AHINE with max chain length $c=1$, with samples containing single relation as training input. (2) The ``underflow'' error happens occasionally, aborting the training process directly. This is due to that the parameters of embedding layer (input layer and softmax layer) change slowly, and the weights of hidden layers change rapidly. We recommend you to give a high learning rate for the embedding layer and give a relatively low rate for the other weights. By doing so, the learning procedure of AHINE will be significantly accelerated and the annoying ``underflow'' error will be fixed.

\subsection{Results in RH}\label{subsection:didi_res}
Node embeddings are learned using the above algorithm and RH graph, in which each node represents a POI grid cell. To evaluate the quality of POI embeddings learned by different methods, several experiments are presented in the following sections. First, we apply them to a real ride-hailing activity prediction in Section ~\ref{subsubsection:act_pred} as there are several kinds of activity predictions, such as demand forecasting ~\cite{demandForcasting}, destination prediction ~\cite{Zhang2017A}, etc. Then in Section ~\ref{subsubsection:poi_vis}, we introduce our POI embedding visualization tool on map and show some interesting results.

\subsubsection{Activity Prediction}\label{subsubsection:act_pred}

We perform one activity prediction experiment on another ride-hailing dataset, which includes $19, 280, 562$ historical orders in Beijing. $246,955$ of these orders are labeled as positive and the rest are labeled as negative.

For each ride-hailing order, we collect a series of related features, including calling time, weather condition, hotel density for departure or destination POI, etc. We call them ``base features''. At the same time, for the POI grid cells where the start and end points are located, we mapped them to the embeddings generated by different models. Finally, we concatenated the features of these two parts together into ``merged features''.

We randomly divide the ride-hailing orders into a training set containing $80\%$ of orders and a test set of  $20\%$. With base features or merged features prepared, we train XGBoost \cite{xgboost} classifiers on the training set and make prediction on the test set. The area under the curve (AUC) is used to measure the performance of final prediction. 

We design two sub tasks of the activity prediction and the results are presented in Table~\ref{tab:drunk}. ``Base'' shows the classifier's performance trained on the base features without POI embeddings. As illustrated in the left part of  Table~\ref{tab:drunk}, we use merged features for classification. As shown in the right part of Table~\ref{tab:drunk}, we make classification directly on POI embeddings without base features. The top $2$ results in each comparison are underlined and the best is marked in bold. 

As shown in Table~\ref{tab:drunk}, among all the models, our proposed method AHINE achieved the best results, while GHINE achieved the second best performance. Moreover, experimental result shows the effectiveness of POI embeddings in the activity prediction, as all the embedding methods are superior than Base. 

% Besides, XGBoost provides an additional feature ranking list. According to the list, we found that POI embeddings do rank in the front position and gain high contribution scores. Among POI embeddings, start POI and dest POI show more importance than calling POI, which consists with our intuition.

\begin{table*}[htbp]
  \caption{Activity prediction results (AUC) in RH data. }
  \label{tab:drunk}
  \centering 
  \begin{tabular}{|c|cc|cc|}
    \hline
    \textbf{Method} & \textbf{AUC (with base)} & \textbf{Gain over base} & \textbf{AUC (without base)} & \textbf{Gain over DailyWalk}\\
    \hline
    Base  & 0.6884 & -      & -      & -      \\
    \hline
    DailyWalk & 0.7285 & 5.82\% & 0.6801 & -      \\
    \hline
    LINE 	  & 0.7484 & 8.72\% & 0.7169 & 5.41\% \\
    \hline
    DeepWalk  & 0.7526 & 9.32\% & 0.7245 & 6.53\% \\
    \hline
    xNetMF    & 0.7480 & 8.66\% & 0.7134 & 4.90\% \\
    \hline
    struc2vec & 0.7401 & 7.51\% & 0.7007 & 3.03\% \\
    \hline
    GHINE     & \underline{0.7594} & \underline{10.31\%} & \underline{0.7316} & \underline{7.57\%}   \\
    \hline
    AHINE     & \underline{\textbf{0.7616}} & \underline{\textbf{10.64\%}} & \underline{\textbf{  0.7327}} & \underline{\textbf{7.73\%}} \\ 
    \hline
  \end{tabular}
\end{table*}

\subsubsection{Visualization}\label{subsubsection:poi_vis}

We build our POI embedding visualization tool based on Baidu Map APIs called ``mapv'' (https://mapv.baidu.com/). We plotted the POIs corresponding to the trained embeddings on the map. In addition, many useful functions are also implemented to explore POIs on the map, such as address-POI translation, POI clustering, similar POI filtering, etc.

We perform the POI clustering experiment in the first place. By applying K-means ($K=20$) \cite{kmeans}, we divide the POIs into $20$ groups and then mark these points in different colors on the Beijing’s map.

\begin{figure*}[ht]
  \centering
  \includegraphics[width=0.6\linewidth]{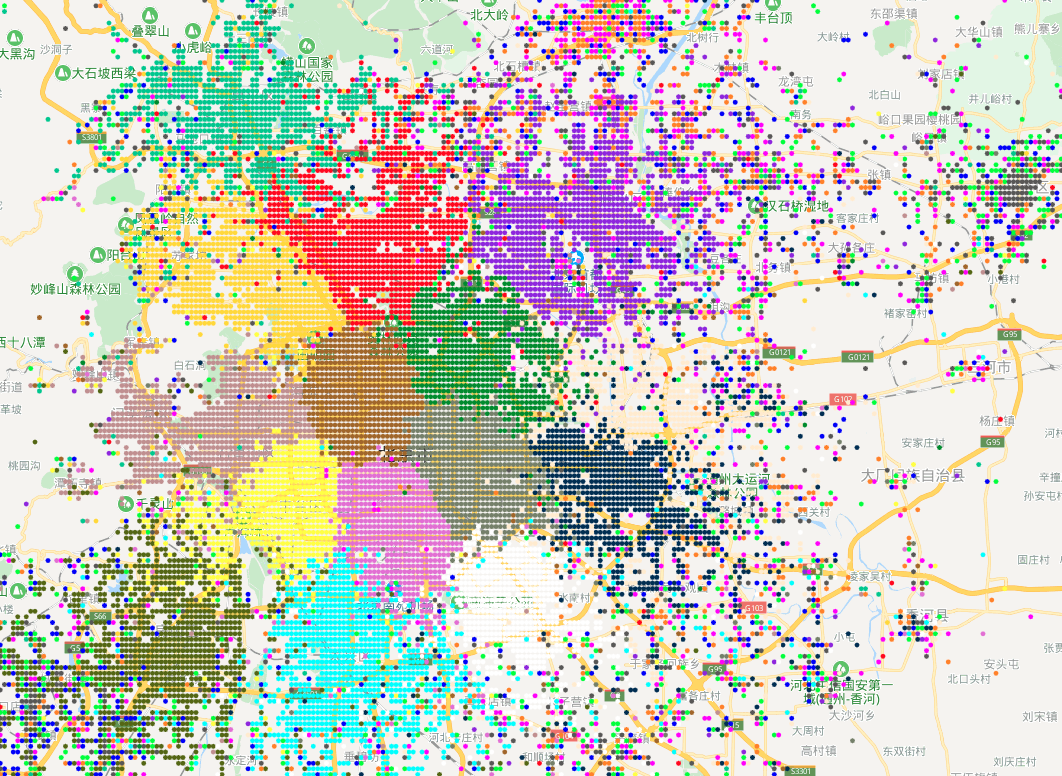}
  \caption{AHINE POI clustering by K-means (K=20) on Beijing's map.}
  \label{fig:clustering}
\end{figure*}

Fig.~\ref{fig:clustering} shows the POI clustering result of AHINE on Beijing's map. Physically closed POIs are usually in the same cluster. It illustrates ability of the model to automatically learn implicitly the geographical proximity relationships between POIs, as during the training process we did not provide any supervised information about geographical location. 

However, location information does not need to be learned because they are usually available, other POI characteristics, such as POI category (e.g., bars, residential areas, business districts, etc.), convenience of transportation facilities, difficulty of parking cars, are much harder to get and represent in form of features. To evaluate if these characteristics are captured by embeddings we can examine top $k$ most similar POI grid cells of unique POI grid cell in the embedding space.

\begin{figure*}[ht]
  \centering
  \includegraphics[width=0.6\linewidth]{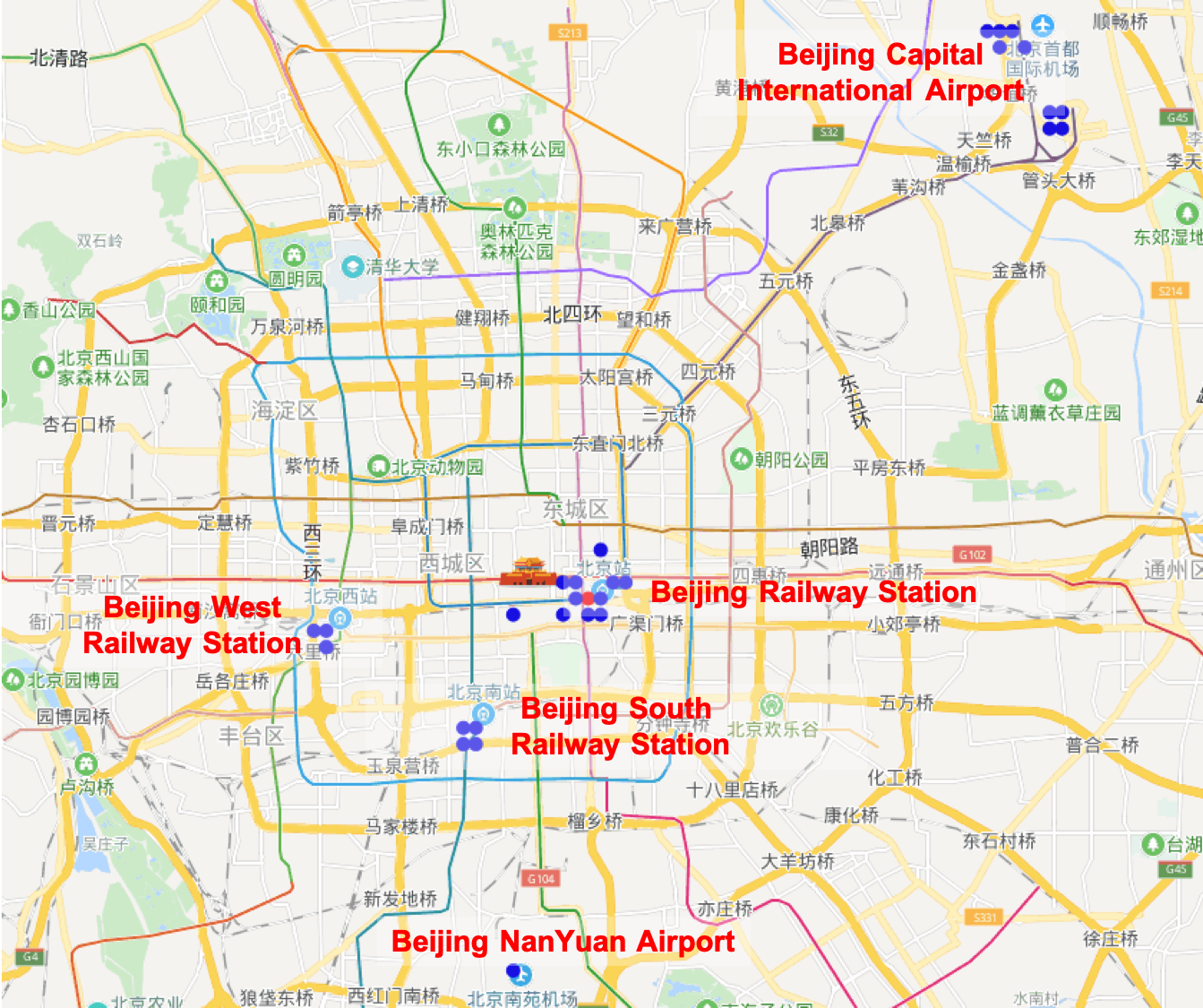}
  \caption{Top 30 similar POIs to Beijing Railway Station.}
  \label{fig:simrank}
\end{figure*}

Fig.~\ref{fig:simrank} shows the top $30$ most similar POIs to Beijing Railway Station. The red point represents the target POI and the blue ones represent the similar POIs. From the map, it is obvious that neighbor POIs do show good similarity. Besides, we surprisingly found other railway stations and airports are also found quite similar to Beijing Railway Station, despite that they are physically far in distance. The result shows that our embeddings are able to capture some semantic relationships (such as type similarity, location proximity, etc.) between nodes in HINs.

\subsection{Results in DBIS \& AMINER}\label{subsection:pub_res}
To label nodes in DBIS and AMINER, we adopt the same third-party labels used in metapath2vec's experiments \cite{metapath2vec}. $8$ categories of venues in Google Scholar are matched with DBIS and AMINER data and we use them to label the corresponding author nodes. In general, we get a valid set with $26,469$ labeled authors and $10$ labeled venues for DBIS. For AMINER, we get $241,235$ labeled authors and $133$ labeled venues. As the number of matched venues is limited, we design three tasks on author including node clustering, node classification and similarity ranking.

\begin{itemize}
\item \textbf{Node Clustering}\\
The learned node embeddings are inputed to a K-means ($K=8$) clustering model. We use normalized mutual information (NMI) \cite{nmi} to evaluate the clustering results.

\item \textbf{Node Classification}\\
In this task, with the embeddings of labeled nodes as input, we try to predict the categories of target authors. A logistic classifier is trained for this eight-class classification task. $80\%$ of the nodes are used for training and the rest $20\%$ for testing. We report the classification performance in terms of both Micro-F1 and Macro-F1 scores \cite{metapath2vec}. 

\item \textbf{Similarity Ranking}\\ % to be rivised
Given a query and returning a ranked item list is one of the main tasks in learning to rank problem \cite{learning2rank}. Mean average precision at K (MAP@K) \cite{MAP} is widely used for evaluating whether a returned list is well ranked or not. In our task, intuitively, a node ought to show high similarities to nodes with the same label and show low similarity to nodes with different labels. We treat a target node as the ``query'' and obtain a returned list by ranking the rest nodes according to their similarity with node ``query''. In the returned list, we define the nodes with the same label as positive samples and the rest are negatives. Thus, we apply the metric MAP@K to our similarity ranking task. 

\end{itemize}

\begin{table}[htbp]
  \caption{Results of three tasks in DBIS data.}
  \label{tab:dbis}
  \centering
  \begin{tabular}{|c|c|cc|c|}
    \hline
    % ~ & Node Clustering & \multicolumn{2}{c|}{Node Classification} & Similarity Ranking \\
    % \midrule
    \textbf{Method} & \textbf{NMI} & \textbf{Macro-F1} & \textbf{Micro-F1} & \textbf{MAP@100} \\
    \hline
    LINE 		  & 0.0918 & 0.2935 & 0.5278 & 0.4023 \\
    \hline
    DeepWalk      & 0.0923 & 0.2687 & 0.5523 & \underline{0.4093} \\
    \hline
    struc2vec    & 0.0561 & 0.1212 & 0.4749 & 0.3530 \\
    \hline
    metapath2vec  & 0.0880 & \underline{\textbf{0.3367}} & \underline{0.5602} & 0.4081 \\
    \hline
    metapath2vec++& 0.0802 & 0.3123 & 0.5416 & 0.3987 \\
    \hline
    HHNE 		  & 0.0753 & 0.2564 & 0.5270 & 0.3875 \\
    \hline
    GHINE         & \underline{\textbf{0.1104}} & 0.3096 & 0.5489 & 0.3921 \\
    \hline
    AHINE         & \underline{0.1100} & \underline{0.3351} & \underline{\textbf{0.5735}} & \underline{\textbf{0.4144}} \\ 
    \hline
  \end{tabular}
\end{table}

The results of author nodes in DBIS dataset is summarized in Table~\ref{tab:dbis}. Same as above, we underline the top $2$ results for each metric and mark the best in bold. As we can observe, the proposed AHINE achieves the best Micro-F1 score in node classification and the best MAP@100 in similarity ranking task. In node clustering task, GHINE and AHINE show similar performance and outperform all baselines significantly. And from GHINE to AHINE, by adding relationship chains, both Macro-F1 and Micro-F1 scores improve greatly.

\begin{table}[htbp]
  \caption{Results of three tasks in AMINER data.}
  \label{tab:aminer}
  \centering
  \begin{tabular}{|c|c|cc|c|}
    \hline
    \textbf{Method} & \textbf{NMI} & \textbf{Macro-F1} & \textbf{Micro-F1} & \textbf{MAP@100} \\
    \hline
    LINE          & 0.5385 & 0.8016 & 0.8191 & 0.6976 \\
    \hline
    DeepWalk      & 0.5306 & 0.8181 & 0.8356 & 0.7188 \\
    \hline
    metapath2vec  & 0.6624 & 0.8655 & 0.8763 & \underline{0.7615} \\
    \hline
    metapath2vec++& 0.5328 & 0.8497 & 0.8617 & 0.7544 \\
    \hline
    HHNE 		  & 0.5330 & 0.8049 & 0.8198 & 0.7319 \\
    \hline
    GHINE         & \underline{\textbf{0.6908}} & \underline{0.8738} & \underline{0.8846} & 0.7356 \\
    \hline
    AHINE         & \underline{0.6816} & \underline{\textbf{0.8786}} & \underline{\textbf{0.8892}} & \underline{\textbf{0.7825}} \\ 
    \hline
  \end{tabular}
\end{table}

Table~\ref{tab:aminer} shows the results of author nodes in AMINER data. AHINE outperforms all the baseline methods in terms of three metrics, and achieves the second best in node clustering task by metric NMI. GHINE also shows satisfactory performance in all tasks except similarity ranking. 

Overall, plenty of experiments are performed in the ride-hailling and bibliograhic datasets. Their results demonstrate the efficiency of proposed GHINE and AHINE models for large-scale HIN embedding.

\section{Conclusion}
In this paper, we propose a general and an adaptive method for the unsupervised embedding learning of heterogeneous information networks. We evaluate the effectiveness of the proposed methods on public datasets. GHINE and AHINE are also applied in a real-world ride-hailing platform to catch the semantic and structure information and help greatly improve the ride-hailing activity prediction service. 

% \section*{Acknowledgment}
% We would like to thank Kevin Knight for his contribution to this project and paper. We would also like to thank the entire DiDi AI Labs NLP and Knowledge Graph team. 

\bibliographystyle{ieeetr}
\bibliography{ahine}
\end{document}